\documentclass[smallextended]{svjour3}       
\smartqed  
\usepackage{graphicx,epsfig}

\journalname{Quantum Inf. Process.}

\begin{document}

\title{
Quantum entanglement in the anisotropic Heisenberg model with multicomponent
DM and KSEA interactions
}

\author{A.~V.~Fedorova \and M.~A.~Yurischev
}


\institute{A.~V.~Fedorova \at 
Institute of Problems of Chemical Physics, Russian Academy of Sciences,
Chernogolovka 142432, Moscow Region, Russia\\     
              \email{panna@icp.ac.ru}           
           \and
M.~A.~Yurischev \at
Institute of Problems of Chemical Physics, Russian Academy of Sciences,
Chernogolovka 142432, Moscow Region, Russia\\     
\email{yur@itp.ac.ru}
}

\date{Received:}

\titlerunning{
Quantum entanglement in the anisotropic Heisenberg model
}
\maketitle

\begin{abstract}
Using group-theoretical approach we found a family of four nine-parameter quantum
states for the two-spin-1/2 Heisenberg system in an external magnetic field and with
multiple components of Dzyaloshinsky-Moriya (DM) and
Kaplan-Shekhtman-Entin-Wohlman-Aharony (KSEA) interactions.
Exact analytical formulas are derived for the entanglement of formation for the
quantum states found.
The influence of DM and KSEA interactions on the behavior of entanglement and on the
shape of disentangled region is studied.
A connection between the two-qubit quantum states and the reduced density matrices of
many-particle systems is discussed.
\end{abstract}


\section{Introduction}
\label{sect:Intro}
Quantum entanglement plays an important role in modern physics.
It is not only used to test some fundamental questions of the quantum mechanics and
quantum information processing but also widely employed in
quantum computing \cite{NC00,VK02,V05,KM06},
communication \cite{E91,W98a,GRTZ02,Pan20},
metrology \cite{PSOST18},
sensing \cite{DRC17,PBGWL18},
imaging \cite{LGB02,ZS20},
simulation of many-body systems \cite{F82,L96,JCJ14,ABC19}.
All of these applications are largely dependent on how to produce entanglement.

Entanglement also affects the thermodynamic behavior of the system
\cite{VA16,BCGAA18,DC19,OSMMN20}.
According to classical thermodynamics, heat engines cyclically operating at a single
temperature are not possible, the maximum efficiency of any heat engine is limited by
the Carnot bound, etc.
In contrast, in quantum thermodynamics, engines that operate at a single temperature
are proposed \cite{SZAW03,G07,YTK17}, entangled thermal systems that can be more
efficient in extracting work than the systems without quantum correlations are
considered \cite{BACRR17,AMVSSOA19,CCA20}, emergence of the second law of
thermodynamics is discussed \cite{LY99,CSHO15,KLSVBL18,BLB19}.

Quantitative measures for the quantum entanglement of bipartite systems were
introduced by Bennett et al. in 1996, first for pure states \cite{BBPS96} and then
for mixed ones \cite{BDSW96}.
These measures are based on entropy and, for mixed states, involve the extremization
procedure to find the optimal ensemble.
Unfortunately, the problems of extemization are known to be very difficult to handle
analytically, and apparently therefore the authors \cite{BDSW96} were able to obtain
an exact expression for the entanglement of formation only in the case of
Bell-diagonal states.
The next important step was taken by Wootters who proved a conjecture of Hill and
Wootters \cite{HW97}, which gives an explicit prescription for evaluating the
entanglement of any two-qubit system.
However, the Wootters formula (see Sect.~\ref{sect:prelim}) requires a solution of
fourth degree algebraic equation, that is, the use of extremely cumbersome Ferrari's
formulas.
Therefore, for practical purposes, it is important to derive compact expressions for
different particular types of quantum states.
For instance, in Ref.~\cite{CW01}, a formula for the two-qubit block-diagonal (XXZ)
state was presented and used.
Then this formula was generalized to the arbitrary X quantum state \cite{YE07}.
In Ref.~\cite{FKY12}, an explicit expression was also obtained for the entanglement
of a centrosymmetric (CS)\footnote{
	A centrosymmetric matrix is a matrix which is symmetric about its center.
	The entries of such a matrix $n\times n$ satisfy the relations
	$a_{i,j}=a_{n+1-i,n+1-j}$ with $i,j=1,\ldots,n$.
	The properties of CS matrices are described in \cite{W85,I93}.
}
quantum state.

Recently, a family of fifteen quantum states was found, for which the problem of
calculating quantum correlations is reduced to solving quadratic equations \cite{Y20}.
This family contains three different types of anisotropic Heisenberg spin systems with
one-component Dzyaloshinsky-Moriya (DM) and Kaplan-Shekhtman-Entin-Wohlman-Aharony
(KSEA) interactions.
In the present paper we give a collection of four two-qubit XYZ models that extend the
DM-KSEA interactions to several components.
A classification was performed by using group-theoretical methods.
The solutions in these cases are reduced to solving cubic algebraic equations.
Using the well-known trigonometric form for the real roots of such equations, we get
closed analytical formulas for the quantum entanglement.

The following should be noted.
A distinctive feature of entanglement in comparison with other quantum correlations,
such as discord, is that in the process of evolution in some parameter (time,
temperature, etc.) it can suddenly disappear \cite{N98,ABV01,YE04}.
This phenomenon is called the entanglement sudden death (ESD) \cite{YE06,YE09}.
Moreover, entanglement can not only suddenly disappear, but also suddenly appear
again \cite{YE09,FT10,WHH18,SG20}.
Below we will present the behavior of quantum entanglement and the shape of the
separability region in the presence of DM and KSEA interactions.

The remainder of the paper is organized as follows.
In the next section (Sect.~\ref{sect:prelim}), we recall the notions and some facts
necessary for further consideration.
Symmetries and corresponding family of quantum states are given in Sect.~\ref{sect:gr-th}. 
Next, in Sect.~\ref{sect:exactEs}, the analytical formulas for the quantum entanglement of the
states found are derived.
The results are discussed in Sect~\ref{sect:Discuss}.
Finally, concluding remarks are provided in Sect.~\ref{sect:Concl}.

\section{
Preliminaries
}
\label{sect:prelim}
The entanglement of formation of a two-qubit quantum state $\rho$
($\rho^\dagger=\rho,\ \rho\ge0,\ {\rm tr}\rho=1$) is given as
\begin{equation}
   \label{eq:E}
   {\rm EoF}(\rho)=-\frac{1+\sqrt{1-C^2}}{2}\log_2\frac{1+\sqrt{1-C^2}}{2}
	 -\frac{1-\sqrt{1-C^2}}{2}\log_2\frac{1-\sqrt{1-C^2}}{2},
\end{equation}
where $C$ is the concurrence.
It is important that the concurrence can also serve as a measure of quantum
correlation.
Wootters \cite{W98} strongly proved that the concurrence of an arbitrary state of two
qubits equals
\begin{equation}
   \label{eq:C}
   C(\rho)=\max\{0,\sqrt{\lambda_1}
	 -\sqrt{\lambda_2}-\sqrt{\lambda_3}-\sqrt{\lambda_4}\},
\end{equation}
where $\lambda_i$ are the eigenvalues (ordered as $\lambda_i\ge\lambda_{i+1}$) of the
$4\times4$-matrix
\begin{equation}
   \label{eq:R}
   R=\rho\tilde\rho,
\end{equation}
in which
\begin{equation}
   \label{eq:tilde_rho}
   \tilde\rho=(\sigma_y\otimes\sigma_y)\rho^*(\sigma_y\otimes\sigma_y)
\end{equation}
is the spin-flipped state.
Here, $\sigma_y$ is the Pauli spin $y$-matrix in standard basis and
the asterisk denotes complex conjugation.

The matrix $\sigma_y\otimes\sigma_y=V$ ($V^{-1}=V$) in Eq.~(\ref{eq:tilde_rho}) acts
as a similarity transformation.
Since $\rho$ is positive operator, the matrix $\tilde\rho$ is also positive and their
product $R$, generally non-Hermitian, will have only real and non-negative
eigenvalues.
Notice in passing the equality $\det R=\det(\rho\tilde\rho)=(\det\rho)^2$.

Further, we will keep in mind that the density matrices come from
the equilibrium statistical mechanics of a system with Hamiltonian $\cal H$, i.e., it
is the Gibbs density matrix
\begin{equation}
   \label{eq:rhoG}
   \rho=\frac{1}{Z}e^{-\beta{\cal H}}
\end{equation}
($Z$ is the partition function and $\beta$ the inverse temperature)
or, say, equals the time-dependent density matrix
\begin{equation}
   \label{eq:rho_t}
   \rho(t)=e^{-i{\cal H}t}\rho(0)e^{i{\cal H}t},
\end{equation}
which follows from the quantum Liouville-von Neumann equation, provided that the
initial state $\rho(0)$ belongs to the same symmetry class as the Hamiltonian.
In these cases, the Hamiltonian and corresponding density matrix commute:
$[\rho,{\cal H}]=0$.
As a result, they can be expanded in terms of the same set of spin operators.

The most general Hamiltonian of two-qubit system can be written as
\begin{equation}
   \label{eq:H}
   {\cal H}={\cal H}_{\rm Z} + {\cal H}_{\rm H} + {\cal H}_{\rm DM} + {\cal H}_{\rm KSEA}.
\end{equation}
Here ${\cal H}_{\rm Z}$ is the Zeeman energy, ${\cal H}_{\rm H}$ denotes the
Heisenberg exchange couplings, the third term
\begin{eqnarray}
   \label{eq:H_DM}
   &&{\cal H}_{\rm DM}={\bf D}\!\cdot\!({\vec\sigma}_1\times{\vec\sigma}_2)
	 \nonumber\\
	 &&=D_x(\sigma_1^y\sigma_2^z-\sigma_1^z\sigma_2^y)
	 +D_y(\sigma_1^z\sigma_2^x-\sigma_1^x\sigma_2^z)
	 +D_z(\sigma_1^x\sigma_2^y-\sigma_1^y\sigma_2^x)
\end{eqnarray}
is the DM interaction, in which ${\bf D}=(D_x,D_y,D_z)$ is the
Dzyaloshinsky vector and $\vec\sigma_i=(\sigma_i^x,\sigma_i^y,\sigma_i^z)$ the
vector of Pauli spin matrices at site $i$~($=1,2$), and the last term
\begin{eqnarray}
   \label{eq:H_KSEA}
	 &&{\cal H}_{\rm KSEA}={\vec\sigma}_1\!\cdot\!\hat{\rm\Gamma}\!\cdot\!{\vec\sigma}_2
	\equiv(\sigma_1^x,\sigma_1^y,\sigma_1^z)
	 \left(
      \begin{array}{ccc}
      .&{\rm\Gamma}_z&{\rm\Gamma}_y\\
      {\rm\Gamma}_z&.&{\rm\Gamma}_x\\
      {\rm\Gamma}_y&{\rm\Gamma}_x&.
      \end{array}
   \right)\!
	 \left(
      \begin{array}{c}
      \sigma_2^x\\
      \sigma_2^y\\
      \sigma_2^z
      \end{array}
   \right)
	 \nonumber\\
	 &&={\rm \Gamma}_x(\sigma_1^y\sigma_2^z+\sigma_1^z\sigma_2^y)
	 +{\rm \Gamma}_y(\sigma_1^z\sigma_2^x+\sigma_1^x\sigma_2^z)
	 +{\rm \Gamma}_z(\sigma_1^x\sigma_2^y+\sigma_1^y\sigma_2^x)
\end{eqnarray}
represents the KSEA interaction, where ${\hat{\rm\Gamma}}$ is a symmetric traceless
tensor; the points are put instead of zero entries.
So, in the general case, both the DM and KSEA interactions have three components each.

\section{
Symmetries and a family of quantum states
}
\label{sect:gr-th}
In open form,
\begin{equation}
   \label{eq:sysy}
   V\equiv\sigma_y\otimes\sigma_y=
	 \left(
      \begin{array}{rrrr}
      .&.&.&-1\\
      .&.&1&.\\
      .&1&.&.\\
      -1&.&.&.
      \end{array}
   \right)=V^T,
\end{equation}
where the subscript $T$ stands for matrix transpose.

The starting point of our approach is as follows.
We will find the symmetries of matrix (\ref{eq:sysy})
and then impose the same symmetry on the state $\rho$.
In this case, the matrix
\begin{equation}
   \label{eq:R1}
   R=\rho\cdot V^T\rho^*V
\end{equation}
is invariant under the taken symmetry.
Further we will use the apparatus of group theory to reduce the matrix $R$ to
block-diagonal form and thereby simplify the problem of extracting its eigenvalues.
Of course, we need to dwell on such symmetries that lead to new results.
For example, matrix (\ref{eq:sysy}) is invariant under transformations
$\sigma_x\otimes\sigma_x$, $\sigma_y\otimes\sigma_y$, and $\sigma_z\otimes\sigma_z$,
but we exclude these symmetries, because they have already been described in
Ref.~\cite{Y20}.

\subsection{
Symmetry groups
}
\label{subsect:P23P14}
It is clear that the matrix (\ref{eq:sysy}) is invariant under simultaneous
permutations of the
second and third rows and columns (transformation $P_{23}$) or, conversely,
under permutations of the first and fourth rows and columns (transformation $P_{14}$).
In an explicit form
\begin{equation}
   \label{eq:P}
   P_{23}=
	 \left(
      \begin{array}{rrrr}
      1&.&.&.\\
      .&.&1&.\\
      .&1&.&.\\
      .&.&.&1
      \end{array}
   \right)=P_{23}^T
\end{equation}
and
\begin{equation}
   \label{eq:P1}
   P_{14}=
	 \left(
      \begin{array}{rrrr}
      .&.&.&1\\
      .&1&.&.\\
      .&.&1&.\\
      1&.&.&.
      \end{array}
   \right)=P_{14}^T.
\end{equation}
Moreover, we also observed that the matrix (\ref{eq:sysy}) remains unchanged under
two other orthogonal transformations
\begin{equation}
   \label{eq:P3}
   P_{\bar2\bar3}=
	 \left(
      \begin{array}{rrrr}
      1&.&.&.\\
      .&.&-1&.\\
      .&-1&.&.\\
      .&.&.&1
      \end{array}
   \right)=P_{\bar2\bar3}^T
\end{equation}
and
\begin{equation}
   \label{eq:P4}
   P_{\bar1\bar4}=
	 \left(
      \begin{array}{rrrr}
      .&.&.&-1\\
      .&1&.&.\\
      .&.&1&.\\
      -1&.&.&.
      \end{array}
   \right)=P_{\bar1\bar4}^T
\end{equation}
Obviously that the squares of matrices (\ref{eq:P})-(\ref{eq:P4}) equal unity.
Each of these four transformations together with the unit element $E$ compose
the groups $\{E,P_{23}\}$, $\{E,P_{14}\}$, $\{E,P_{\bar2\bar3}\}$, and
$\{E,P_{\bar1\bar4}\}$.
Below we sometimes use the collective notation
$P=\{P_{23},P_{14},P_{\bar2\bar3},P_{\bar1\bar4}\}$ for the permutation
transformations.

\subsection{
Group-theoretical analysis
}
\label{subsect:g-ta}
To know simplifications for the initial task due to symmetries, we perform a
group-theoretical analysis.
Each of groups $\{E,P\}$ has the second order and has two irreducible representations
${\rm\Gamma}^{(1)}$ and ${\rm\Gamma}^{(2)}$.
The $4\times4$ unit matrix and any matrix $P$ together give the original
representation $\rm\Gamma$ of the taken group in the space of density matrix.
The characters of $\rm\Gamma$ (traces of representation matrices) equal $\chi(E)=4$
and $\chi(P_{23})=\chi(P_{14})=\chi(P_{\bar2\bar3})=\chi(P_{\bar1\bar4})=2$.
Knowing them we can find the multiplicities $a_1$ and $a_2$ with which the irreducible
representations ${\rm\Gamma}^{(1)}$ and ${\rm\Gamma}^{(2)}$, respectively,
are contained in $\rm\Gamma$.
\begin{table}[t]
\caption{
Character table of the group $\{E,P\}$
\upshape\upshape}
\label{tab:chi}
\begin{tabular}{lcc}
\hline\noalign{\smallskip}
$\{E,P\}$ & $E$ & $P$ \\
\noalign{\smallskip}\hline\noalign{\smallskip}
${\rm\Gamma}^{(1)}$ & $1$ & $1$ \\
\smallskip
${\rm\Gamma}^{(2)}$ & $1$ & $-1\ \ $ \\
$\rm\Gamma$ & $4$ & $2$ \\
\noalign{\smallskip}\hline
\end{tabular}
\end{table}
For this purpose it is sufficient to use of the character table for the group
(Table~\ref{tab:chi}) and the formula \cite{LL_QM65,H62}
\begin{equation}
   \label{eq:a_mu}
   a_\mu=\frac{1}{g}\sum_G\chi(G)\chi^{(\mu)^*}(G),
\end{equation}
where $g$ is the order of the group, $\chi^{(\mu)}(G)$ the character of the element
$G$ in the $\mu$-th irreducible representation, and $\chi(G)$ the character of the
same element in the representation under question.
Simple calculations yield
\begin{equation}
   \label{eq:a1a2}
   a_1=3,\qquad a_2=1.
\end{equation}
This means that in the basis where the representation $\rm\Gamma$ of the Abelian group
$\{E,P\}$ is completely reduced, the density matrices will take a block-diagonal form
with one subblock $3\times3$ and one ``subblock'' $1\times1$.

The eigenvalues
of the matrices $P$ are equal to $+1$ which is threefold degenerate
and to $-1$.
The eigenvectors are
\begin{equation}
   \label{eq:v1-4}
   |\Phi^+\rangle=
   \frac{1}{\sqrt2}
	 \left(
      \begin{array}{c}
      1\\
      .\\
      .\\
      1
      \end{array}
   \right),
   |\Phi^-\rangle=
   \frac{1}{\sqrt2}
	 \left(
      \begin{array}{c}
      1\\
      .\\
      .\\
      -1
      \end{array}
   \right),
   |\Psi^+\rangle=
   \frac{1}{\sqrt2}
	 \left(
      \begin{array}{c}
      .\\
      1\\
      1\\
      .
      \end{array}
   \right),
   |\Psi^-\rangle=
   \frac{1}{\sqrt2}
	 \left(
      \begin{array}{c}
      .\\
      1\\
      -1\\
      .
      \end{array}
   \right)
\end{equation}
(these are the Bell states).
The given eigenvectors are the same for all $P$-operators (and also for $V$), but the
eigenvectors which corresponds to the non-degenerate eigenvalue $-1$ are different for
each $P$-matrix.
Vector $|\Psi^-\rangle$ in set~(\ref{eq:v1-4}) corresponds to the operator $P_{23}$,
$|\Phi^-\rangle$ to $P_{14}$, $|\Psi^+\rangle$ to $P_{\bar2\bar3}$, and 
$|\Phi^+\rangle$ to $P_{\bar1\bar4}$.

Quasi-diagonalizing transformations are constructed from the eigenvectors of the
operators $P$.
Such a transformation for the operator $P_{23}$ can be written as
\begin{equation}
   \label{eq:O1}
   O_{23}=\frac{1}{\sqrt{2}}
	 \left(
      \begin{array}{ccrr}
      .&1&1&.\\
      1&.&.&1\\
      1&.&.&-1\\
      .&1&-1&.
      \end{array}
   \right)=O_{23}^T.
\end{equation}
This transformation is orthogonal and symmetric.
Likewise for other operators $P$:
\begin{equation}
   \label{eq:O2}
   O_{14}=\frac{1}{\sqrt{2}}
	 \left(
      \begin{array}{ccrr}
      1&.&.&1\\
      .&1&1&.\\
      .&1&-1&.\\
      1&.&.&-1
      \end{array}
   \right)=O_{14}^T,
\end{equation}
\begin{equation}
   \label{eq:O3}
   O_{\bar2\bar3}=\frac{1}{\sqrt{2}}
	 \left(
      \begin{array}{ccrr}
      .&1&1&.\\
      1&.&.&1\\
      -1&.&.&1\\
      .&1&-1&.
      \end{array}
   \right),
\end{equation}
and
\begin{equation}
   \label{eq:O4}
   O_{\bar1\bar4}=\frac{1}{\sqrt{2}}
	 \left(
      \begin{array}{ccrr}
      1&.&.&1\\
      .&1&1&.\\
      .&1&-1&.\\
      -1&.&.&1
      \end{array}
   \right).
\end{equation}
The last columns in the matrices (\ref{eq:O1})--(\ref{eq:O4}) equal the eigenvectors
corresponding the eigenvalue $-1$.
Such an arrangement of columns provides a direct sum structure $3\oplus1$ for the
density matrices after their quasi-diagonalizations.

Let us now turn to the consideration of models with four symmetries separately.

\subsection{
Construction of quantum states
}
\label{subsect:q-st}
The most general matrix which commutes with the operator (\ref{eq:P}) has the
following form
\begin{equation}
   \label{eq:A}
   A=
	 \left(
      \begin{array}{cccc}
      A_1&A_2&A_2&A_3\\
      A_4&A_5&A_6&A_7\\
      A_4&A_6&A_5&A_7\\
      A_8&A_9&A_9&A_{10}
      \end{array}
   \right),
\end{equation}
where $A_1,\ldots,A_{10}$ are arbitrary real or complex numbers.
This matrix remains unchanged while permuting the second and third rows and columns
(for brevity, it can be called a $\pi_{23}$-matrix).

Since the density matrix must be Hermitian,
the most general form of density  matrix that commutes with $P_{23}$ is written as
\begin{equation}
   \label{eq:rhoP}
   \rho_{23}=
	 \left(
      \begin{array}{cccc}
      a&\mu&\mu&\nu\\
      \mu^*&b&c&\gamma\\
      \mu^*&c&b&\gamma\\
      \nu^*&\gamma^*&\gamma^*&d
      \end{array}
   \right).
\end{equation}
Here and below, the Latin letters $a$,
$b$, $c$, and $d$ for the entries in density matrices are real, whereas the Greek
letters $\mu$, $\nu$, and $\gamma$ can be complex.
Since for any density matrix ${\rm tr}\rho=1$, the matrix (\ref{eq:rhoP}) contains
nine real parameters.

Bloch decomposition of the density matrix (\ref{eq:rhoP}) is written as
\begin{eqnarray}
   \label{eq:rhoP_Bloch}
   &&\rho_{23}=\frac{1}{4}[1+s_x(\sigma_1^x+\sigma_2^x)+s_y(\sigma_1^y+\sigma_2^y)+s_z(\sigma_1^z+\sigma_2^z)
   +c_1\sigma_1^x\sigma_2^x+c_2\sigma_1^y\sigma_2^y
	 \nonumber\\
   &&+c_3\sigma_1^z\sigma_2^z+g_x(\sigma_1^y\sigma_2^z+\sigma_1^z\sigma_2^y)+g_y(\sigma_1^x\sigma_2^z+\sigma_1^z\sigma_2^x)
	 +g_z(\sigma_1^x\sigma_2^y+\sigma_1^y\sigma_2^x)].
\end{eqnarray}
This matrix in open form looks as
\begin{equation}
   \label{eq:rhoP_open}
   \rho_{23}=\frac{1}{4}
	 \left(
      \begin{array}{cccc}
      1+2s_z+c_3&s_x+g_y&s_x+g_y&c_1-c_2-2ig_z\\
      &-i(s_y+g_x)&\ -i(s_y+g_x)&\\ \\
			s_x+g_y&1-c_3&c_1+c_2&s_x-g_y\\
      +i(s_y+g_x)&&&-i(s_y-g_x)\\ \\
			s_x+g_y&c_1+c_2&1-c_3&s_x-g_y\\
      +i(s_y+g_x)&&&-i(s_y-g_x)\\ \\
			c_1-c_2+2ig_z&s_x-g_y&s_x-g_y&1-2s_z+c_3\\
      &+i(s_y-g_x)&\ +i(s_y-g_x)&
			\end{array}
   \right).
\end{equation}
Comparing Eqs.~(\ref{eq:rhoP}) and (\ref{eq:rhoP_open}) one find relation between
parameters of the quantum state $\rho_{23}$ in different forms,
\begin{eqnarray}
   \label{eq:param_scg}
   &&a=(1+2s_z+c_3)/4,\quad b=(1-c_3)/4,\quad c=(c_1+c_2)/4,
   \nonumber\\
   &&d=(1-2s_z+c_3)/4,\quad \mu=(s_x+g_y)/4-i(s_y+g_x)/4,\\
   &&\nu=(c_1-c_2)/4-ig_z/2,\quad \gamma=(s_x-g_y)/4-i(s_y-g_x)/4
   \nonumber.
\end{eqnarray}
Here, nine real parameters $s_x$, $s_y$, $s_z$, $c_1$, $c_2$, $c_3$, $g_x$, $g_y$, and
$g_z$ equal unary and binary correlation functions and can vary from $-1$ to $+1$.
The Hamiltonian with the same algebraic structure reads
\begin{equation}
   \label{eq:H23}
   {\cal H}_{23}={\bf B}\!\cdot\!(\vec\sigma_1
	 +\vec\sigma_2)+J_x\sigma_1^x\sigma_2^x+J_y\sigma_1^y\sigma_2^y+J_z\sigma_1^z\sigma_2^z
   +{\vec\sigma}_1\!\cdot\!\hat{\rm\Gamma}\!\cdot\!{\vec\sigma}_2,
\end{equation}
where ${\bf B}=(B_x, B_y, B_z)$ is an external magnetic field with arbitrary
orientation, $J_\alpha$ ($\alpha=x, y, z$) are the Heisenberg coupling constants,
and the last term represents the complete KSEA interactions.

Acting in a similar manner, we obtain density matrices and corresponding
Hamiltonians for the systems with other three symmetries.
Commutativity condition of Hermitian matrix with the $P_{14}$ leads to the $\pi_{14}$
quantum state:
\begin{eqnarray}
   \label{eq:rhoP1}
   \rho_{14}=
	 \left(
      \begin{array}{cccc}
      a&\mu&\nu&d\\
      \mu^*&b&\gamma&\mu^*\\
      \nu^*&\gamma^*&c&\nu^*\\
      d&\mu&\nu&a
      \end{array}
   \right)=
	 &&\frac{1}{4}[1+s_x(\sigma_1^x+\sigma_2^x)+s_y(\sigma_1^y-\sigma_2^y)
	 +s_z(\sigma_1^z-\sigma_2^z)
	 \nonumber\\
   &&+c_1\sigma_1^x\sigma_2^x+c_2\sigma_1^y\sigma_2^y
   +c_3\sigma_1^z\sigma_2^z+g_x(\sigma_1^y\sigma_2^z+\sigma_1^z\sigma_2^y)
	 \nonumber\\
	 &&+\delta_y(\sigma_1^z\sigma_2^x-\sigma_1^x\sigma_2^z)
	 +\delta_z(\sigma_1^x\sigma_2^y-\sigma_1^y\sigma_2^x)],
\end{eqnarray}
where $s_x$, $s_y$, $s_z$, $c_1$, $c_2$, $c_3$, $g_x$, $\delta_y$, and $\delta_z$ are
nine real parameters; they also have the physical meaning of correlation functions.
The Hamiltonian of the system is written as
\begin{eqnarray}
   \label{eq:H14}
   &&{\cal H}_{14}=B_x(\sigma_1^x+\sigma_2^x)+B_y(\sigma_1^y-\sigma_2^y)
	 +B_z(\sigma_1^z-\sigma_2^z)
   +J_x\sigma_1^x\sigma_2^x+J_y\sigma_1^y\sigma_2^y
	 \nonumber\\
   &&+J_z\sigma_1^z\sigma_2^z+{\rm\Gamma}_x(\sigma_1^y\sigma_2^z+\sigma_1^z\sigma_2^y)
	 +D_y(\sigma_1^x\sigma_2^z-\sigma_1^z\sigma_2^x)
	 +D_z(\sigma_1^x\sigma_2^y-\sigma_1^y\sigma_2^x)].\quad\ 
\end{eqnarray}
For the system with $P_{\bar2\bar3}$-symmetry the density matrix is given by
\begin{eqnarray}
   \label{eq:rhoP3_Bloch}
   &&\rho_{\bar2\bar3}=\frac{1}{4}[1+s_x(\sigma_1^x-\sigma_2^x)+s_y(\sigma_1^y-\sigma_2^y)
	 +s_z(\sigma_1^z+\sigma_2^z)
   +c_1\sigma_1^x\sigma_2^x+c_2\sigma_1^y\sigma_2^y
	 \nonumber\\
   &&+c_3\sigma_1^z\sigma_2^z+\delta_x(\sigma_1^y\sigma_2^z-\sigma_1^z\sigma_2^y)
	 +\delta_y(\sigma_1^z\sigma_2^x-\sigma_1^x\sigma_2^z)
	 +g_z(\sigma_1^x\sigma_2^y+\sigma_1^y\sigma_2^x)]
\end{eqnarray}
and corresponding Hamiltonian is
\begin{eqnarray}
   \label{eq:H23bar}
   &&{\cal H}_{\bar2\bar3}=B_x(\sigma_1^x-\sigma_2^x)+B_y(\sigma_1^y-\sigma_2^y)
	 +B_z(\sigma_1^z+\sigma_2^z)
   +J_x\sigma_1^x\sigma_2^x+J_y\sigma_1^y\sigma_2^y
	 \nonumber\\
   &&+J_z\sigma_1^z\sigma_2^z+D_x(\sigma_1^y\sigma_2^z-\sigma_1^z\sigma_2^y)
	 +D_y(\sigma_1^x\sigma_2^z-\sigma_1^z\sigma_2^x)
	 +{\rm\Gamma}_z(\sigma_1^x\sigma_2^y-\sigma_1^y\sigma_2^x)].\quad\ 
\end{eqnarray}
Finally, for the system with $P_{\bar1\bar4}$-symmetry quantum state is written as
\begin{eqnarray}
   \label{eq:rhoP4_Bloch}
   &&\rho_{\bar1\bar4}=\frac{1}{4}[1+s_x(\sigma_1^x-\sigma_2^x)+s_y(\sigma_1^y+\sigma_2^y)
	 +s_z(\sigma_1^z-\sigma_2^z)
   +c_1\sigma_1^x\sigma_2^x+c_2\sigma_1^y\sigma_2^y
	 \nonumber\\
   &&+c_3\sigma_1^z\sigma_2^z+\delta_x(\sigma_1^y\sigma_2^z-\sigma_1^z\sigma_2^y)
	 +g_y(\sigma_1^z\sigma_2^x+\sigma_1^x\sigma_2^z)
	 +\delta_z(\sigma_1^x\sigma_2^y-\sigma_1^y\sigma_2^x)]
\end{eqnarray}
and the Hamiltonian equals
\begin{eqnarray}
   \label{eq:H14bar}
   &&{\cal H}_{\bar1\bar4}=B_x(\sigma_1^x-\sigma_2^x)+B_y(\sigma_1^y+\sigma_2^y)
	 +B_z(\sigma_1^z-\sigma_2^z)
   +J_x\sigma_1^x\sigma_2^x+J_y\sigma_1^y\sigma_2^y
	 \nonumber\\
   &&+J_z\sigma_1^z\sigma_2^z+D_x(\sigma_1^y\sigma_2^z-\sigma_1^z\sigma_2^y)
	 +{\rm\Gamma}_y(\sigma_1^x\sigma_2^z+\sigma_1^z\sigma_2^x)
	 +D_z(\sigma_1^x\sigma_2^y-\sigma_1^y\sigma_2^x)].\quad\ 
\end{eqnarray}
The family of four quantum states found is presented in Table~\ref{tab:4qs}.
%
\begin{table}[t]
\caption{
Four quantum states and Hamiltonians generated by $P$-operators and, in
braces, the sets of nine combinations of the Pauli spin operators required for their
Bloch decompositions.
Entries $a$, $b$, $c$, and $d$ are real and $\mu$, $\nu$, and $\gamma$ can be complex
\upshape\upshape}
\label{tab:4qs}
\begin{tabular}{cc}
\noalign{\smallskip}\hline
\hline\noalign{\smallskip}
\smallskip
$P_{23}$ & $P_{14}$\\
\smallskip
${\cal H}_{23}, \rho_{23}\!\!:$&${\cal H}_{14}, \rho_{14}\!\!:$\\
\smallskip
$
\left(
   \begin{array}{cccc}
    a&\mu&\mu&\nu\\
    \mu^*&b&c&\gamma\\
    \mu^*&c&b&\gamma\\
    \nu^*&\gamma^*&\gamma^*&d
    \end{array}
\right)
$&$
\left(
   \begin{array}{cccc}
    a&\mu&\nu&d\\
    \mu^*&b&\gamma&\mu^*\\
    \nu^*&\gamma^*&c&\nu^*\\
    d&\mu&\nu&a
    \end{array}
\right)
$\\
\smallskip
\{$\sigma_1^x+\sigma_2^x,\ \sigma_1^y+\sigma_2^y,\ \sigma_1^z+\sigma_2^z$,&
\{$\sigma_1^x+\sigma_2^x,\ \sigma_1^y-\sigma_2^y,\ \sigma_1^z-\sigma_2^z$,\\
\smallskip
$\sigma_1^x\sigma_2^x$,\ $\sigma_1^y\sigma_2^y$,\ $\sigma_1^z\sigma_2^z,$&
$\sigma_1^x\sigma_2^x$,\ $\sigma_1^y\sigma_2^y$,\ $\sigma_1^z\sigma_2^z,$\\
\smallskip
$\sigma_1^y\sigma_2^z+\sigma_1^z\sigma_2^y$,&$\sigma_1^y\sigma_2^z+\sigma_1^z\sigma_2^y$,\\
\smallskip
$\sigma_1^z\sigma_2^x+\sigma_1^x\sigma_2^z$,&$\sigma_1^z\sigma_2^x-\sigma_1^x\sigma_2^z$,\\
\smallskip
$\sigma_1^x\sigma_2^y+\sigma_1^y\sigma_2^x$\}&$\sigma_1^x\sigma_2^y-\sigma_1^y\sigma_2^x$\}\\
\noalign{\smallskip}\hline\noalign{\smallskip}
\smallskip
$P_{\,\bar2\bar3}$ & $P_{\,\bar1\bar4}$\\
\smallskip
${\cal H}_{\bar2\bar3}, \rho_{\bar2\bar3}\!\!:$&${\cal H}_{\bar1\bar4}, \rho_{\bar1\bar4}\!\!:$\\
\smallskip
$
\left(
   \begin{array}{cccc}
    a&\mu&-\mu&\nu\\
    \mu^*&b&c&\gamma\\
    -\mu^*&c&b&-\gamma\\
    \nu^*&\gamma^*&-\gamma^*&d
    \end{array}
\right)
$&$
\left(
   \begin{array}{cccc}
    a&\mu&\nu&d\\
    \mu^*&b&\gamma&-\mu^*\\
    \nu^*&\gamma^*&c&-\nu^*\\
    d&-\mu&-\nu&a
    \end{array}
\right)
$\\
\smallskip
\{$\sigma_1^x-\sigma_2^x,\ \sigma_1^y-\sigma_2^y,\ \sigma_1^z+\sigma_2^z$,&
\{$\sigma_1^x-\sigma_2^x,\ \sigma_1^y+\sigma_2^y,\ \sigma_1^z-\sigma_2^z$,\\
\smallskip
$\sigma_1^x\sigma_2^x$,\ $\sigma_1^y\sigma_2^y$,\ $\sigma_1^z\sigma_2^z,$&
$\sigma_1^x\sigma_2^x$,\ $\sigma_1^y\sigma_2^y$,\ $\sigma_1^z\sigma_2^z,$\\
\smallskip
$\sigma_1^y\sigma_2^z-\sigma_1^z\sigma_2^y$,&$\sigma_1^y\sigma_2^z-\sigma_1^z\sigma_2^y$,\\
\smallskip
$\sigma_1^z\sigma_2^x-\sigma_1^x\sigma_2^z$,&$\sigma_1^z\sigma_2^x+\sigma_1^x\sigma_2^z$,\\
\smallskip
$\sigma_1^x\sigma_2^y+\sigma_1^y\sigma_2^x$\}&$\sigma_1^x\sigma_2^y-\sigma_1^y\sigma_2^x$\}\\
\noalign{\smallskip}\hline
\hline
\end{tabular}
\end{table}
%
In this table, we also wrote out the spin matrices required for the Bloch expansions
of each quantum state and corresponding Hamiltonian.

Notice that the matrices having the structure of each of the found density operators
are algebraically closed: their sums and products preserve the same structure.
At the same time, they, generally speaking, do not commute with each other.

In conclusion of this section we note the following.
These models first of all allow for fully anisotropic Heisenberg couplings $J_x$,
$J_y$, and $J_z$.
An external magnetic field may also be present, but this is different for different
models.
If in the model with $P_{23}$-symmetry, Eq.~(\ref{eq:H23}), the external field is
uniform, then in the other three systems the field is parallel for one spin component
and antiparallel for the other two components [see Eqs~(\ref{eq:H14}),
(\ref{eq:H23bar}), and (\ref{eq:H14bar})].
Further, in all models, only three components (out of six possible $D_x$, $D_y$,
$D_z$, ${\rm\Gamma}_x$, ${\rm\Gamma}_y$, and ${\rm\Gamma}_z$) of DM-KSEA interactions
can be present in total.
We see that the Hamiltonian with $P_{23}$ symmetry includes a complete set of KSEA
bonds, while the DM interactions are completely absent.
In the other three models, the three mixed components (two DM and one KSEA) of the
DM-KSEA interactions are distributed in such a way that each component of the KSEA
interaction has a similar component of a parallel external field [see again
Eqs~(\ref{eq:H14}), (\ref{eq:H23bar}), and (\ref{eq:H14bar})].
Park \cite{P19} recently discussed a two-qubit XXZ model with two DM components, but
in the absence of an external magnetic field and without the KSEA interaction.

The analytical formulas for the quantum entanglement of the four states presented in
Table~\ref{tab:4qs} are considered in the next section.

\section{
Exact formulas for the quantum entanglement
}
\label{sect:exactEs}
The density matrix $\rho_{23}$ in a quasidiagonal representation is written as
\begin{eqnarray}
   \label{eq:OrhoOa}
   &&\rho_{23}^\prime=O_{23}^T\rho_{23}O_{23}
	 =\left(
	 \begin{array}{cc}
   r&.\\
	 .&(1-c_1-c_2-c_3)/4
   \end{array}
	 \right)\\
	 &&\equiv\frac{1}{4}\left(
      \begin{array}{cccc}
      1+c_1+c_2-c_3&2s_x+2ig_x&2g_y+2is_y&.\\
      2s_x-2ig_x&1+c_1-c_2+c_3&2s_z+2ig_z&.\\
      2g_y-2is_y&2s_z-2ig_z&1-c_1+c_2+c_3&.\\
      .&.&.&1-c_1-c_2-c_3
      \end{array}
   \right).
	 \nonumber
\end{eqnarray}
The domain of definition ${\cal D}\subseteq[-1,1]^9$ for the model parameters is
defined, in accord with Sylvester-like criterion, by four conditions for the main
minors of first order
\begin{eqnarray}
   \label{eq:M1}
   &&1-c_1+c_2+c_3\ge0,\quad 1+c_1-c_2+c_3\ge0,\quad 1+c_1+c_2-c_3\ge0,
	 \nonumber\\
	 &&\quad 1-c_1-c_2-c_3\ge0,
\end{eqnarray}
by three conditions for the minors of second order
\begin{eqnarray}
   \label{eq:M2}
   &&(1-c_1+c_2+c_3)(1+c_1-c_2+c_3)-4(s_z^2+g_z^2)\ge0,
	 \nonumber\\
   &&(1+c_1-c_2+c_3)(1+c_1+c_2-c_3)-4(s_x^2+g_x^2)\ge0,\\
   &&(1-c_1+c_2+c_3)(1+c_1+c_2-c_3)-4(s_y^2+g_y^2)\ge0,
	 \nonumber
\end{eqnarray}
and by one condition for the minor of third order
\begin{eqnarray}
   \label{eq:M3}
   &&(1-c_1+c_2+c_3)(1+c_1-c_2+c_3)(1+c_1+c_2-c_3)
	 \nonumber\\
	 &&-4[(1-c_1+c_2+c_3)(s_x^2+g_x^2)+(1+c_1-c_2+c_3)(s_y^2+g_y^2)\\
	 &&+(1+c_1+c_2-c_3)(s_z^2+g_z^2)]
   +16(g_xs_ys_z+g_ys_xs_z+g_zs_xs_y-g_xg_yg_z)\ge0.
	 \nonumber
\end{eqnarray}
Thus, the body ${\cal D}$ is bounded by four planes, three surfaces of second
order, and one cubic surface.

The spin-flipped state $\tilde\rho_{23}^\prime=O_{23}^T\tilde\rho O_{23}$  has similar
block-diagonal structure.
Hence the matrix $R^\prime=\rho^\prime\tilde\rho^\prime$ also has the quasidiagonal
form:
\begin{equation}
   \label{eq:КP_qu}
   R^\prime=
	 \left(
      \begin{array}{cc}
      Q&.\\
      .&(1-c_1-c_2-c_3)^2/16
      \end{array}
   \right).
\end{equation}
Thus, one eigenvalue
\begin{equation}
   \label{eq:lam4aa}
   \lambda_4=(1-c_1-c_2-c_3)^2/16
\end{equation}
of the matrix $R$ has already been found.
The remaining $3\times3$-subblock $Q$ has entries $Q_{ij}=q_{ij}/16$ with
\begin{eqnarray}
   \label{eq:qij}
   &&q_{11}=\alpha_3^2+4(g_x^2+g_y^2-s_x^2-s_y^2)-8i(g_xs_x-g_ys_y),
	 \nonumber\\
   &&q_{12}=2s_x(\alpha_2-\alpha_3)+4(g_zs_y-g_ys_z)+2i[g_x(\alpha_2+\alpha_3)-2(g_yg_z+s_ys_z)],
	 \nonumber\\
   &&q_{13}=2g_y(\alpha_1+\alpha_3)-4(g_xg_z+s_xs_z)+2i[s_y(\alpha_1-\alpha_3)+2(g_zs_x-g_xs_z)],
	 \nonumber\\
   &&q_{21}=2s_x(\alpha_3-\alpha_2)-4(g_zs_y-g_ys_z)-2i[g_x(\alpha_2+\alpha_3)-2(g_yg_z+s_ys_z)],
	 \nonumber\\
   &&q_{22}=\alpha_2^2+4(g_x^2+g_z^2-s_x^2-s_z^2)+8i(g_xs_x-g_zs_z),\\
   &&q_{23}=2s_z(\alpha_1-\alpha_2)+4(g_ys_x-g_xs_y)+2i[g_z(\alpha_1+\alpha_2)-2(g_xg_y+s_xs_y)],
	 \nonumber\\
   &&q_{31}=2g_y(\alpha_1+\alpha_3)-4(g_xg_z+s_xs_z)+2i[s_y(\alpha_1-\alpha_3)+2(g_zs_x-g_xs_z)],
	 \nonumber\\
   &&q_{32}=2s_z(\alpha_2-\alpha_1)-4(g_ys_x-g_xs_y)-2i[g_z(\alpha_1+\alpha_2)-2(g_xg_y+s_xs_y)],
	 \nonumber\\
   &&q_{33}=\alpha_1^2+4(g_y^2+g_z^2-s_y^2-s_z^2)-8i(g_ys_y-g_zs_z),
	 \nonumber
\end{eqnarray}
where
\begin{equation}
   \label{eq:a1a2a3}
   \alpha_1=1-c_1+c_2+c_3,\quad \alpha_2=1+c_1-c_2+c_3,\quad \alpha_3=1+c_1+c_2-c_3.
\end{equation}
Using these expressions we find three invariants of matrix $Q$ which are
needed to obtain the corresponding secular equation.
The trace of $Q$ equals
\begin{equation}
   \label{eq:TrQaa}
   {\rm Tr}\,Q=\frac{1}{16}[\alpha_1^2+\alpha_2^2+\alpha_3^2+8(g_x^2+g_y^2+g_z^2-s_x^2-s_y^2-s_z^2)].
\end{equation}
The sum of main minors of the second order,
$m={\rm Tr}\,Q^{-1}\det Q=({\rm Tr}^2Q-{\rm Tr}\,Q^2)/2$,
is expressed via the model parameters as
\begin{eqnarray}
   \label{eq:sum_minors-2aa}
	 &&m=\frac{1}{16^2}\{\alpha_1^2\alpha_2^2+\alpha_2^2\alpha_3^2+\alpha_1^2\alpha_3^2
	 +8[\alpha_1^2(g_x^2-s_x^2)+\alpha_2^2(g_y^2-s_y^2)+\alpha_3^2(g_z^2-s_z^2)
	 \nonumber\\
	 &&-\alpha_1\alpha_2(g_z^2+s_z^2)-\alpha_2\alpha_3(g_x^2+s_x^2)-\alpha_1\alpha_3(g_y^2+s_y^2)]
   +16[(g_x^2+g_y^2+g_z^2)^2
	 \nonumber\\
	 &&+(s_x^2+s_y^2+s_z^2)^2]
	 +32[g_xg_yg_z(\alpha_1+\alpha_2+\alpha_3)+g_xs_ys_z(-\alpha_1+\alpha_2+\alpha_3)\\
	 &&+g_ys_xs_z(\alpha_1-\alpha_2+\alpha_3)+g_zs_xs_y(\alpha_1+\alpha_2-\alpha_3)-s_x^2(-g_x^2+g_y^2+g_z^2)-s_y^2(g_x^2
	 \nonumber\\
	 &&-g_y^2+g_z^2)-s_z^2(g_x^2+g_y^2-g_z^2)]
	 -128(g_xg_ys_xs_y+g_yg_zs_ys_z+g_xg_zs_xs_z)\}.
	 \nonumber
\end{eqnarray}
Finally, the determinant $\det Q=\det^2r$ is given as
\begin{eqnarray}
   \label{eq:detQaa}
   &&\det Q=\frac{1}{16^3}\{\alpha_1\alpha_2a_3-4[\alpha_1(g_x^2+s_x^2)+\alpha_2(g_y^2+s_y^2)+\alpha_3(g_z^2+s_z^2)]
	 \nonumber\\
   &&+16(g_xs_ys_z+g_ys_xs_z+g_zs_xs_y-g_xg_yg_z)\}^2.
\end{eqnarray}
All these three invariants are real.

The secular equation of $3\times3$ subblock $Q$ is given as
\begin{equation}
   \label{eq:sec_eq_Qa}
   \lambda^3-\lambda^2{\rm tr}\,Q+m\lambda-\det Q=0.
\end{equation}
Since the eigenvalues of matrix $R$ are real, it is suitable to use the trigonometric
form for the roots of cubic equation \cite{KK,M57}.
As a result,
\begin{eqnarray}
   \label{eq:lam123aa}
	 &&\lambda_k=\frac{1}{3}{\rm tr}\,Q
	 +\frac{2}{3}({\rm tr}^2Q
	 \nonumber\\
	 &&-3m)^{1/2}\cos\!\bigg[\frac{1}{3}\arccos\bigg(\frac{2{\rm tr}^3Q-9m{\rm tr}Q
	 +27\det Q}{2({\rm tr}^2Q-3m)^{3/2}}\bigg)
	 +\frac{2\pi}{3}(k-1)\bigg],\ 
\end{eqnarray}
where $k=1,2,3$.
These three eigenvalues together with the fourth one (\ref{eq:lam4aa})
and Eqs.~(\ref{eq:E}) and (\ref{eq:C}) give a closed analytical formula for
the quantum entanglement of two-qubit system with $P_{23}$-symmetry.

Note the following.
Particular calculations show that the largest eigenvalue of $R$ is really $\lambda_1$
from the set (\ref{eq:lam123aa}).
Nevertheless, for reliability, one should sort the eigenvalues, find the largest among
them and assign it the designation $\lambda_1$.

Further, the states $\rho_{14}$, $\rho_{\bar2\bar3}$, and $\rho_{\bar1\bar4}$ are
reduced to block-diagonal forms $3\oplus1$ using orthogonal transformations
$O_{14}$, $O_{\bar2\bar3}$, and $O_{\bar1\bar4}$ respectively.
Therefore, the corresponding $R$-matrices have such a block-diagonal structure.
As a result, we obtain in a similar way exact analytical expressions for the quantum
entanglement for the named three states

\section{
Discussion
}
\label{sect:Discuss}
We have obtained a family of exactly solvable models with several components of the
DM-KSEA interactions.
This allows us to perform investigation and compare the behavior of quantum
entanglement in them.

\subsection{
Impact of KSEA and DM interactions on entanglement
}
\label{subsect:Compare}
The model with $P_{23}$-symmetry has three components of KSEA interaction.
Each of other three models has one component of KSEA interaction and two components of
DM interaction.
The difference between these three models consists in different distribution of KSEA
and DM components.
Therefore to compare a role KSEA interaction in the model with $P_{23}$-symmetry and
DM interaction in other three models it is enough to take one model, say, with
$P_{14}$-symmetry.

\begin{figure}[t]
\begin{center}
\epsfig{file=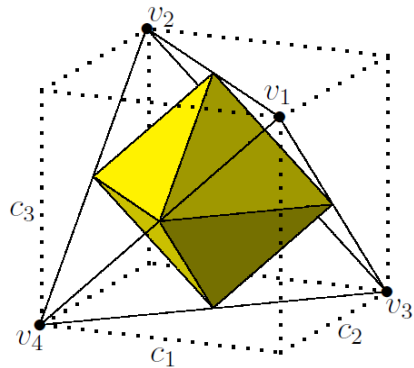,width=6.8cm}
\end{center}
\begin{center}
\caption{(Color online)
Tetrahedron ${\cal T}$ with vertices $v_1$, $v_2$, $v_3$, and $v_4$ is the domain for
the Bell-diagonal states.
The yellow octahedron ${\cal O}$ is the set of separable states.
The remaining part ${\cal T}\setminus{\cal O}$ consists of four small tetrahedra,
where the entanglement is nonzero
}
\label{fig:z_octa}
\end{center}
\end{figure}

The domain of definition for its nine parameters $s_x$, $s_y$, $s_z$, $c_1$, $c_2$,
$c_3$, $g_x$, $g_y$, and $g_z$ of model with $P_{23}$-symmetry is defined by
conditions (\ref{eq:M1})-(\ref{eq:M3}).
In the limit of Bell-diagonal states, $s_x=s_y=s_z=g_x=g_y=g_z=0$, the domain for the
parameters $(c_1, c_2, c_3)$ is reduced to the tetrahedron $\cal T$  in which the
separability region is confined to the inscribed octahedron $\cal O$ specified as
$|c_1|+|c_2|+|c_3|\le1/2$ \cite{HH96} (see Fig.~\ref{fig:z_octa}).

\begin{figure}[t]
\begin{center}
\epsfig{file=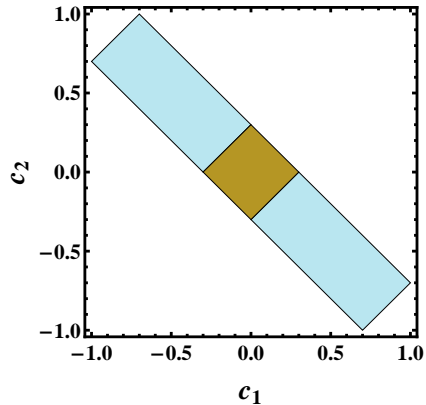,width=5.6cm}
\caption{(Color online)
Section of tetrahedron $\cal T$ and octahedron ${\cal O}$ (see
Fig.~\ref{fig:z_octa}) by plane $c_3=0.7$.
The filled rectangle is the domain for the arguments $c_1$ and $c_2$.
The blue areas of the rectangle correspond to non-zero quantum entanglement, and the
square yellow area (section of octahedron ${\cal O}$) answers to the zero entanglement
}
\label{fig:z07bd}
\end{center}
\end{figure}

For a clear visualization of the picture, we take slices of a three-dimensional
space $(c_1, c_2, c_3)$ by planes $c_3 = const$.
In this case, the cross section by the plane $c_3 = 0.7$ for Bell-diagonal states
looks like shown in fig.~\ref{fig:z07bd}.

Let us examine the changes that occur in the form of separability regions and in
behavior of entanglement when two components of the KSEA interaction in the
$\rho_{23}$-model are replaced by two analogous components of the DM interaction in
the $\rho_{14}$-model: $(g_y,g_z)\to(\delta_y,\delta_z)$;
in other words, when the $\rho_{23}$-model transforms into the $\rho_{14}$-model
while maintaining the same strengths of interaction constants. 

The domain of definitions for the nine parameters
$s_x$, $s_y$, $s_z$, $c_1$, $c_2$, $c_3$, $g_x$, $\delta_y$, and $\delta_z$ of the
$\rho_{14}$ state in its quasidiagonal form is defined by the inequalities on the main
minors:
by four conditions for the minors of first order which coincide with (\ref{eq:M1}) and
hence with the rectangle domain for the Bell-diagonal states shown in
Fig~\ref{fig:z07bd},
by three conditions for the minors of second order
\begin{eqnarray}
   \label{eq:M2a}
   &&(1+c_1+c_2-c_3)(1+c_1-c_2+c_3)-4(s_x^2+g_x^2)\ge0,
	 \nonumber\\
   &&(1-c_1-c_2-c_3)(1+c_1-c_2+c_3)-4(s_y^2+\delta_y^2)\ge0,\\
   &&(1-c_1-c_2-c_3)(1+c_1+c_2-c_3)-4(s_z^2+\delta_z^2)\ge0,
	 \nonumber
\end{eqnarray}
and by one condition for the minor of third order
\begin{eqnarray}
   \label{eq:M3a}
   &&(1-c_1-c_2-c_3)(1+c_1+c_2-c_3)(1+c_1-c_2+c_3)
	 \nonumber\\
	 &&-4[(1-c_1-c_2-c_3)(s_x^2+g_x^2)+(1+c_1+c_2-c_3)(s_y^2+\delta_y^2)\\
	 &&+(1+c_1-c_2+c_3)(s_z^2+\delta_z^2)]
   -16(s_xs_y\delta_z+g_xs_ys_z+g_x\delta_y\delta_z-s_x\delta_ys_z)\ge0.
	 \nonumber
\end{eqnarray}
If the parameters $s_x$, $s_y$, $s_z$, $g_x$, $g_y$, $g_z$, $\delta_x$, $\delta_y$,
and $\delta_z$ begin to change, then the tetrahedron $\cal T$ and region ${\cal O}$
will deform.
We found the region of separable states using the positive partial transpose (PPT)
criterion (the Peres-Horodecki criterion) \cite{HH96,P96,HHHH09}.
For this, an equation $\lambda_{min}=0$ was numerically solved, where $\lambda_{min}$
is the lowest eigenvalue of the partially transposed density matrix.

\begin{figure}[t]
\begin{center}
\epsfig{file=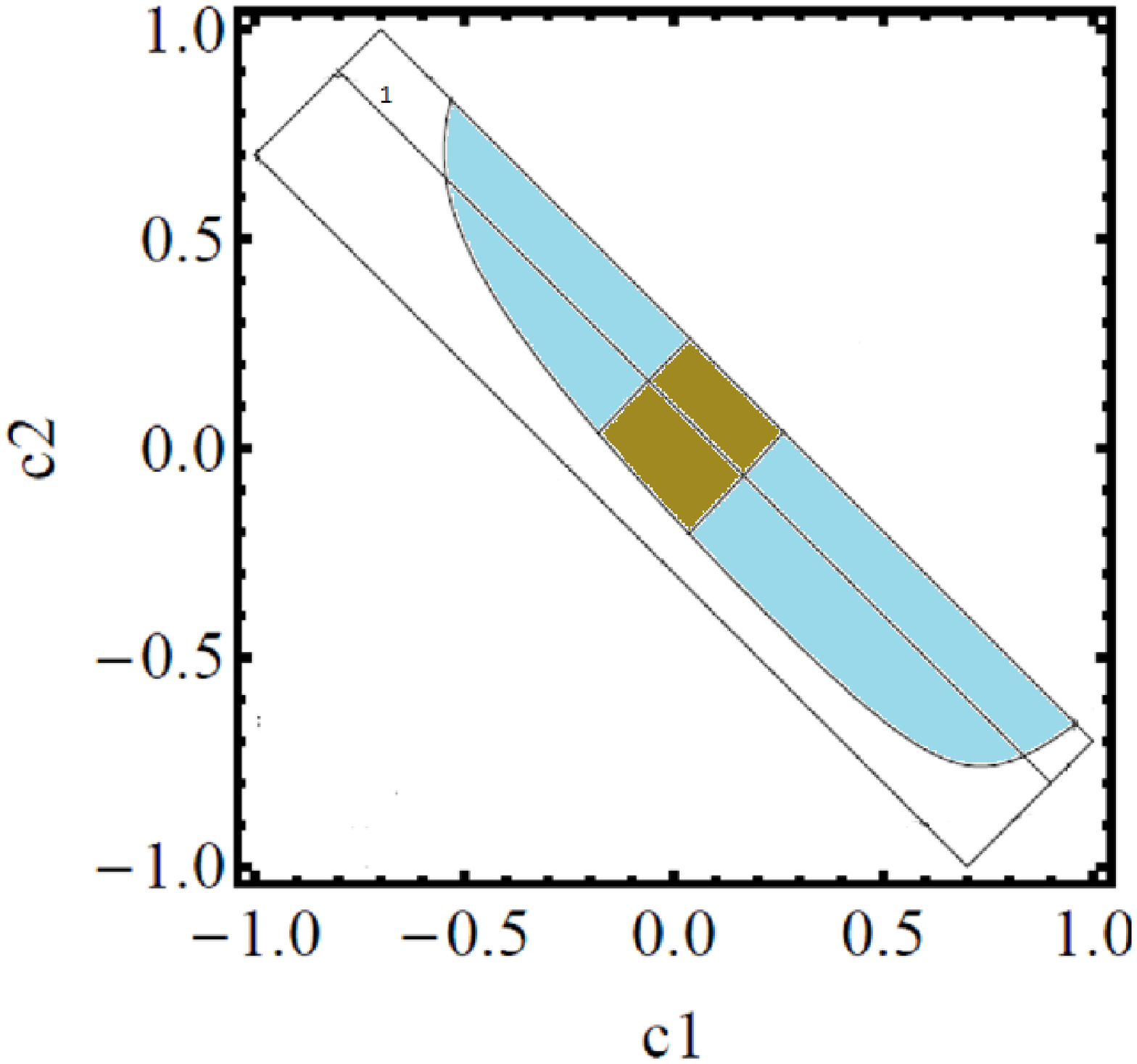,width=5.6cm}
\hspace{5mm}
\epsfig{file=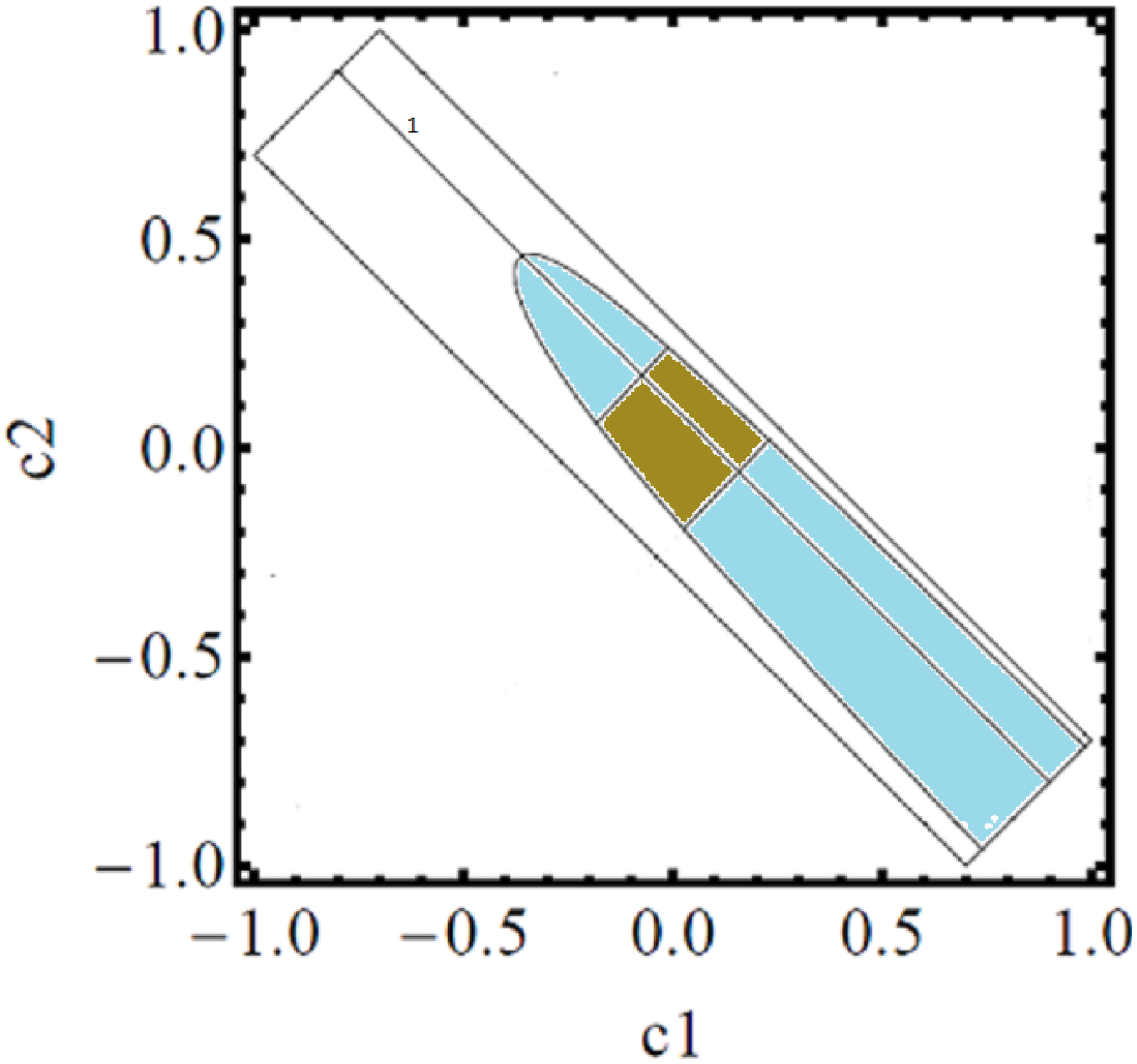,width=5.6cm}
\caption{(Color online)
Phase diagrams for the states $\rho_{23}$ (left) and  $\rho_{14}$ (right) by
$c_3=0.7$, $s_x=0.15$, $s_y=s_z=0$, $g_x=0.16$,
$g_y=\delta_y=0.1$, and $g_z=\delta_z=0.04$.
The filled parts of rectangles are the domains of definition.
The blue areas correspond to non-zero quantum entanglement while the
yellow ones to zero entanglement.
Straight lines 1 are the paths $c_2=-c_1+0.1$ along which the quantum entanglement
is evaluated
}
\label{fig:z07_23-14bc}
\end{center}
\end{figure}

Figure~\ref{fig:z07_23-14bc} shows how the domains and separability regions change
when two components of KSEA interaction replaced by analogous components of DM
interaction.
It is seen that the structure of regions modifies very significantly.

\begin{figure}[t]
\begin{center}
\epsfig{file=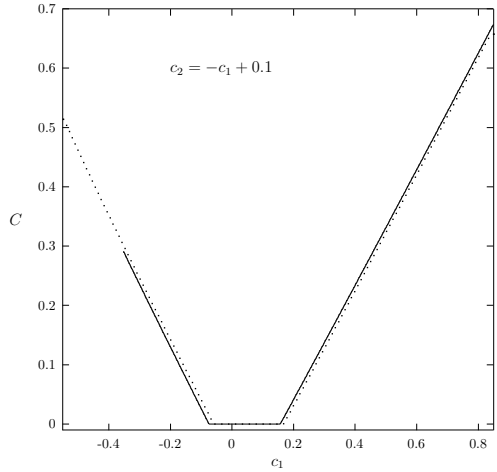,width=6.4cm}
\caption{
Quantum concurrence $C$ vs $c_1$ along the path $c_2=-c_1+0.1$
by $c_3=0.7$, $s_x=0.15$, $s_y=s_z=0$, $g_x=0.16$, $g_y=\delta_y=0.1$, and
$g_z=\delta_z=0.04$.
The dotted line corresponds to the purely KSEA interactions while the solid curve
corresponds to the replacement $g_y$ and $g_z$ components by the $\delta_y$ and
$\delta_z$ ones with the same strengths
}
\label{fig:zc2-c1+01a}
\end{center}
\end{figure}

The behavior of quantum entanglement in both cases under discussion is presented in
Fig.~\ref{fig:zc2-c1+01a}.
We observe that the behavior of quantum entanglements in both cases, on the contrary,
changes little on the whole.
When the parameter $c_1$ begins to increase, the entanglements first decrease
monotonously (see Fig.~\ref{fig:zc2-c1+01a}).
At $c_1\approx-0.06$, the sudden death of entanglement happens and the concurrence
curve has a fracture.
The systems then go through the death zone where entanglement absent.
However, when the control parameter $c_1$ reaches the critical value $c_1\approx0.16$,
the entanglements reincarnate and start to grow as shown in Fig.~\ref{fig:zc2-c1+01a}.

Notice the following.
If the phase diagram is known, the schemes for paths can be chosen so to avoid
sudden death of entanglement or, conversely, to kill him, if required (e.g., for
switches of quantum entanglement).

\subsection{
Two-qubit states as reduced density matrices of many-body systems
}
\label{subsect:many-body}
If there is a $N$-particle system in the state $\rho_N$, one can obtain two-particle
correlations between arbitrary two particles $i$ and $j$ using the reduced density
matrix
\begin{equation}
   \label{eq:rho_ij}
   \rho_{ij}={\rm tr}_{\{1,2,\ldots,N\}\setminus\{ij\}}\rho_N.
\end{equation}
This circumstance makes it important to search for various families of two-site
quantum states for which quantum correlations can be estimated \cite{XZYL08}.

It is known \cite{CW01,W02,ON02,GMPPR20} that the two-qubit reduced density matrix
of one-dimensional XY chain has block-diagonal or X structures.
Further, it was shown in \cite{FKY12} that the reduced density matrix of the
nuclear spin pairs in a gas of $N$ molecules or atoms (for example, $^{129}{\rm Xe}$)
in closed nanopore which is in a strong magnetic field has the CS structure.
All these quantum states are part of a collection of fifteen quantum states that was
recently described in Ref.~\cite{Y20}.

Consider now $N$-qubit system which is symmetric under any permutations of the
particles, i.e., under transformations of the symmetric group $S_N$.
A large number of experimentally relevant multi-atom states exhibit such a symmetry.
The symmetric $N$-qubit quantum state up to terms of second order can be written as
\begin{equation}
   \label{eq:rho_N-sym}
   \rho_N^{sym}=\frac{1}{2^N}\biggl[1+\sum_{\alpha=x,y,z}s_\alpha\sum_{i=1}^N\sigma_i^\alpha
	  +\sum_{\alpha,\beta=x,y,z}t_{\alpha\beta}\sum_{i<j}\sigma_i^\alpha\sigma_j^\beta
		+O(\sigma\sigma\sigma)\biggr],
\end{equation}
where $s_\alpha$ are the components of average normalized spins of the qubits
($s_\alpha=\langle\sigma_i^\alpha\rangle$, $\forall i=1,\ldots,N$) and
$t_{\alpha\beta}=t_{\beta\alpha}$ are the elements of the real, symmetric two-qubit
correlation matrix that, without loss of generality, can be written as
\begin{equation}
   \label{eq:hat-t}
   \hat t=
	 \left(
      \begin{array}{ccc}
      c_1&g_z&g_y\\
      g_z&c_2&g_x\\
      g_y&g_x&c_3
      \end{array}
   \right).
\end{equation}
Then the reduced density matrix for a pair of qubits is written as
\begin{equation}
   \label{eq:rho-sym_12}
   \rho_2^{sym}={\rm tr}_{\{3,\ldots,N\}}\rho_N^{sym}
	 =\frac{1}{4}[1+{\bf s}(\vec\sigma_1+\vec\sigma_2)+\vec\sigma_1\cdot\hat t\cdot\vec\sigma_2]
\end{equation}
which exactly coincides with the density matrix $\rho_{23}$ given by
Eq.~(\ref{eq:rhoP_Bloch}).
This is not surprising, because the expansion of the operator $P_{23}$ in
spin Pauli matrices has the form
\begin{equation}
   \label{eq:P-Bl}
   P_{23}\equiv
	 \left(
      \begin{array}{rrrr}
      1&.&.&.\\
      .&.&1&.\\
      .&1&.&.\\
      .&.&.&1
      \end{array}
   \right)=\frac{1}{2}(1+\vec\sigma_1\vec\sigma_2).
\end{equation}
The right hand side of this equality represents the well-known Dirac spin exchange
operator \cite{D29,D30}.
Hence, the two-qubit reduced density matrix of symmetric state has the
structure that defined by Eq.~(\ref{eq:rhoP}).
This result was first established in \cite{WM02} (see also \cite{WS03,V06,KD19}). 
In our paper we generalized these calculations, because we obtained an analytical
formula for the pair entanglement that is suitable for the multi-qubit symmetric
systems.

\section{Concluding remarks and outlook}
\label{sect:Concl}
In this paper, we have found the class of four quantum states for the two-qubit
Heisenberg model in an external magnetic field and with multiple components of both
antisymmetric Dzyaloshinsky-Moriya and symmetric
Kaplan-Shekhtman-Entin-Wohlman-Aharony interactions.
The states are collected in Table~\ref{tab:4qs}.
Then we have derived closed analytical formulas for the quantum entanglement of the
found states.
Thus, an assortment of families of quantum states for which quantum
correlations can be obtained in closed analytical forms is extended.

Moreover, we have investigated the effect of DM-KSEA interactions on the phase
diagrams of entangled-disentangled states and on the behavior of quantum entanglement.
Two-qubit quantum states, for which it is possible to get explicit expressions for the
quantum correlations, are important both in themselves and as reduced density matrices
of many-particle systems.

\vspace{-10mm}
\section*{}
{\bf Acknowledgment}\ 
This work was performed as a part of the the program CITIS \# AAAA-A19-119071190017-7.



\end{document}